\RequirePackage{fix-cm}
\documentclass[smallextended]{svjour3}       
\smartqed  
\usepackage{graphicx}
\usepackage{epsfig}
\usepackage{graphicx}        
\usepackage{float}
\usepackage[round]{natbib}

\usepackage[english]{babel}
\usepackage{amssymb,amstext,amsmath}
\usepackage{multicol}
\usepackage[table]{xcolor}

\begin{document}

\title{On the consistency of a spatial-type interval-valued median for random intervals
}


\author{Beatriz Sinova        \and
        Stefan Van Aelst 
}

\authorrunning{B. Sinova, S. Van Aelst} 

\institute{B. Sinova \at
              Departamento de Estad\'{\i}stica e I.O. y D.M., \\ Universidad de Oviedo, 33071 Oviedo, Spain \\
              \email{sinovabeatriz@uniovi.es}           
           \and
           B. Sinova \and S. Van Aelst  \at
              Department of Applied Mathematics, Computer Science and Statistics, \\ Ghent University, 9000 Gent, Belgium
           \and
           S. Van Aelst \at
           Department of Mathematics, KU Leuven, 3001 Leuven, Belgium
}

\date{Received: date / Accepted: date}

\maketitle

\begin{abstract}
The sample $d_\theta$-median is a robust estimator of the central tendency or location of an interval-valued random variable. While the interval-valued sample mean can be highly influenced by outliers, this spatial-type interval-valued median remains much more reliable. 
In this paper, we show that under general conditions the sample $d_\theta$-median is a strongly consistent estimator of the $d_\theta$-median of an interval-valued  random variable.
\end{abstract}

\section{Introduction}
\label{author sec:1}
In this data driven area, the amount and complexity of the available data grows at an almost incredible speed. Therefore, there is a high need to develop novel tools to cope with such complex data structures. Whereas the first statistical techniques were designed only to manage either quantitative or qualitative data, we can now find statistical procedures to handle functional data (see for instance Arribas-Gil and M\"{u}ller~\citeyear{Arribas2014}; Febrero-Bande and Gonz\'{a}lez-Manteiga~\citeyear{Febrero2013}; Jacques and Preda~\citeyear{jacques}), fuzzy-valued data (see, for instance, Ferraro and Giordani~\citeyear{Ferraro2013}; Gonz\'{a}lez-Rodr\'{\i}guez \emph{et al.}~\citeyear{Gonzalez2012}; Coppi \emph{et al.}~\citeyear{Coppi2012}); incomplete/missing data (see, for instance, Bianco \emph{et al.}~\citeyear{bianco}; Ferraty \emph{et al.}~\citeyear{ferraty}; Lin~\citeyear{lin}; Zhao \emph{et al.}~\citeyear{zhao}), and several other types of data.

Interval-valued data are a type of complex data that requires specific statistical techniques to analyze them. Interval-valued data may arise for different reasons. In some cases the underlying random variable is intrinsically interval-valued, e.g. the daily fluctuation of the systolic blood pressure.
In other cases, there is an underlying real-valued but to preserve a level of confidentiality respondents are only asked to indicate the interval containing their value, e.g. their salary.
It may also happen that the real-valued measurement is only partially known due to certain limitations, such as is the case for interval censored data.
Finally, aggregation of a typically large dataset may lead to e.g. interval-valued symbolic data which include interval variation and structure.

The $d_\theta$-median considered here does not make any assumption about the source of the interval-valued data. In particular, it does not matter whether the random experiment that generated the data involves an underlying observable real-valued random variable or not.
An important remark is that the space of intervals is only a semilinear space, but not a linear space due to the lack of the opposite of an interval.  
Therefore, although intervals can be identified with two-dimensional vectors (with first component the mid-point/centre and second component the nonnegative spread/radius), it is not advisable to treat them as regular bivariate data. Indeed, common assumptions for multivariate techniques do not hold in this case.

Statistical procedures for random interval-valued data have already been proposed in the literature for different purposes, such as regression analysis (e.g Gil \emph{et al.}~\citeyear{gil2002,Gil2007}; Gonz\'{a}lez-Rodr\'{\i}guez \emph{et al.}~\citeyear{Gonzalez2012}; Blanco-Fern\'{a}ndez \emph{et al.}~\citeyear{Blanco2011,Blanco2013}; Lima Neto \emph{et al.}~\citeyear{Lima2011}; Fagundes \emph{et al.}~\citeyear{fagundes2013}; Giordani~\citeyear{Giordani2014}); testing hypotheses (e.g. Montenegro \emph{et al.}~\citeyear{Montenegro2008}; Nakama \emph{et al.}~\citeyear{Nakama2010}; Gonz\'{a}lez-Rodr\'{\i}guez \emph{et al.}~\citeyear{Gonzalez2012}), clustering (e.g. De Carvalho \emph{et al.}~\citeyear{Carvalho2006}; D'Urso \emph{et al.}~\citeyear{DUrso2006,DUrso2011,DUrso2014}; Giusti and Grassini~\citeyear{Giusti2008}; Da Costa \emph{et al.}~\citeyear{DaCosta2013}, etc.), principal component analysis (e.g.  Billard and Diday~\citeyear{Billard2003}; D'Urso and Giordani~\citeyear{DUrso2004}; Makosso-Kallyth and Diday~\citeyear{Makosso2012}, etc.), modelling distributions (see Brito and Duarte Silva~\citeyear{Brito2012}; Sun and Ralescu~\citeyear{SunRalescu2014}).

One of the most commonly used location measures is the Aumann-type mean (see Aumann \citeyear{Aumann1965}). It is indeed supported by numerous valuable properties, including laws of Large Numbers, and is also coherent with the interval arithmetic. The main disadvantage is that it is strongly influenced by outliers and data changes, which makes this measure not always suitable as a summary measure of the distribution of a random interval. This drawback is in fact inherited from the standard real/vectorial-valued case. In the real case, the most popular alternative is the median.

In the real case, the most popular robust alternative to the mean is the median.
For multivariate data the spatial median (also called the $L_1$-median, as introduced by Weber~\citeyear{Weber1909}) is a popular robust alternative to estimate the center of the multivariate data.
The spatial median is defined as the point in multivariate space with minimal average Euclidean distance to the observations. For more details and extensions, see for instance Gower~(\citeyear{gower1974}), Brown~(\citeyear{Brown}), Milasevic and Ducharme~(\citeyear{ducharme1987}),  (Cadre \citeyear{cadre2001}), Roelant and Van Aelst~(\citeyear{Roelant}), Debruyne \emph{et al.}~(\citeyear{Debruyne}), Fritz \emph{et al.}~(\citeyear{Fritz}), Zuo~(\citeyear{Zuo}).

Sinova and Van Aelst (2014) adapted the spatial median to interval-valued data by using a suitable $L^2$ metric on this space (see also Sinova et al. \citeyear{sinova2013}). They used the versatile generalized metric introduced by Bertoluzza \emph{et al.} (\citeyear{bertoluzza1995}, see also Gil \emph{et al.}~\citeyear{gil2002}; Trutschnig \emph{et al.}~\citeyear{trutschnig2009}) 
The resulting $d_\theta$-median estimator has been shown to be robust with high breakdown point and good finite-sample properties.  In this paper we show another important property of the estimator, which is its strong consistency.

The rest of this paper is organized as follows: in Section 2 the basic concepts related to the interval-valued space, interval arithmetic and metric for intervals will be introduced, as well as the usual location measure. In Section 3, the $d_\theta$-median for random intervals and its main properties are recalled. The strong consistency of the $d_\theta$-median is proven in Section 4. Finally, some concluding remarks are presented in Section 5.

\section{The $d_\theta$-median of a random interval}
\label{author_sec:3}

Let $\mathcal K_c(\mathbb R)$ denote the class of nonempty compact intervals. Any interval $K$ in the space $K_c(\mathbb R)$ can be characterized in terms of either its infimum and supremum, $K = [\inf K, \sup K]$, or its mid-point and spread or radius, $K=[{\hbox {\rm mid}}\, K-{\hbox {\rm spr}}\, K, {\hbox {\rm mid}}\, K +{\hbox {\rm spr}}\, K]$, where
$${\hbox {\rm mid}}\, K = \frac{\inf K + \sup K}{2}, \quad {\hbox {\rm spr}}\, K = \frac{\sup K - \inf K}{2} \geq 0.\;$$
The usual interval arithmetic provides the addition, i.e. $K + K' = [\inf K + \inf K', \sup K+\sup K']$
with $K,K'\in\mathcal K_c(\mathbb R)$ and the product by a scalar, i.e.
$\gamma\cdot K = [\gamma\cdot \mathrm{mid}\,K - |\gamma|\cdot \mathrm{spr}\,K,\gamma\cdot \mathrm{mid}\,K + |\gamma|\cdot \mathrm{spr}\,K]$ with $K\in\mathcal K_c(\mathbb R)$ and 
$\gamma \in \mathbb R$.
With these two operations the space $\mathcal K_c(\mathbb R)$ is semilinear, but not linear due to the lack of a difference of intervals. Therefore, statistical techniques for interval-valued data are based on distances.

To measure the distance between two interval-valued observations, we consider the \emph{$d_\theta$ metric} introduced by Bertoluzza \emph{et al.}~(\citeyear{bertoluzza1995}), which can be defined as (see Gil \emph{et al.}~\citeyear{gil2002}):
$$d_\theta(K,K')=\sqrt{({\hbox {\rm mid}}\, K-{\hbox {\rm mid}}\, K')^2+\theta\cdot({\hbox {\rm spr}}\, K-{\hbox {\rm spr}}\, K')^2},$$
where $K,K'\in \mathcal K_c(\mathbb R)$ and $\theta\in(0,\infty)$. 
Following the general random set approach, a
\emph{random interval} can usually defined as a Borel measurable mapping $X:\Omega\rightarrow\mathcal K_c(\mathbb R)$, where $(\Omega,\mathcal A,P)$ is a probability space with respect to $\mathcal A$ and on $K_c(\mathbb R)$ the Borel $\sigma$-field generated by the topology induced by
the $d_\theta$ metric. As a consequence from the Borel measurability, crucial concepts in probabilistic and inferential developments, such as the (induced) distribution of a random interval or the stochastic independence of random intervals, are well-defined.

One of the most used location measures is the \emph{Aumann-type mean value}. It is defined, if it exists, as the interval $E[X]=[E(\inf X),E(\sup X)]$ or $E[X]=[E({\hbox {\rm mid}}\, X)-E({\hbox {\rm spr}}\, X),E({\hbox {\rm mid}}\, X)+E({\hbox {\rm spr}}\, X)]$ (both expressions are equivalent). Moreover, it is the Fr\'echet expectation with respect to the $d_\theta$ metric, i.e., it is the unique interval that minimizes, over $K\in\mathcal K_c(\mathbb R)$, the expression $E[(d_\theta(X,K))^2]$. 

As a robust alternative to the Aumann-type mean, Sinova and Van Aelst (2014) proposed the $d_\theta$-median as measure of location, which is defined as follows.
\begin{definition} The \emph{$d_{\theta}$-median(s)} of a random interval $X:\Omega\rightarrow \mathcal K_c(\mathbb R)$ is(are) the interval(s) $\mathrm{M}_\theta[X]\in \mathcal K_c(\mathbb R)$ such that
$$E(d_{\theta}(X,\mathrm{M}_\theta[X])) = \min_{K\in\mathcal K_c(\mathbb R)}E(d_{\theta}(X,K)),$$
whenever the involved expectations exist.
\end{definition}
%
Analogously, the sample $d_\theta$-median statistic is defined as follows.
\begin{definition}
Let $(X_1,\ldots,X_n)$ be a simple random sample from a random interval $X:\Omega\rightarrow \mathcal K_c(\mathbb R)$ with realizations $\mathbf{x}_n=(x_1,\ldots,x_n)$. The \emph{sample $d_{\theta}$-median} (or medians) $\widehat{\mathrm{M}_\theta[X]}_n$ is (are) the random interval that takes, for $\mathbf{x}_n$, the interval value(s) $\widehat{\mathrm{M}[\mathbf{x}_n]}$ that is (are) the solution(s) of the following optimization problem: 

$$\begin{array}{l}\displaystyle{\min_{K\in\mathcal K_c(\mathbb R)}} \frac{1}{n}\sum_{i=1}^n d_{\theta}(x_i,K)
\\ \displaystyle{=\min_{(y,z)\in \mathbb R\times [0,\infty)}}\frac{1}{n}\sum_{i=1}^n \sqrt{({\hbox {\rm mid}}\, x_i-y)^2+\theta\cdot({\hbox {\rm spr}}\, x_i-z)^2}\end{array}$$

\noindent where $K$, $y$ and $z$ depend on $\mathbf{x}_n$ (which has been omitted from the notation for the sake of simplicity) and the fixed value $\theta$.
\end{definition}

Sinova and Van Aelst (2014) showed the existence of the sample $d_\theta$-median estimator and its uniqueness whenever not all the two-dimensional sample points $\{({\hbox {\rm mid}}\, x_i,{\hbox {\rm spr}}\, x_i)\}_{i=1}^n$ are collinear. Moreover, the robustness was shown by its 
finite sample breakdown point (Donoho and Huber~\citeyear{donoho1983}) which is given by $$\mathrm{fsbp}\big(\widehat{\mathrm{M}_\theta[X]}_n,\mathbf{x}_n,d_\theta\big)=\frac{1}{n}\cdot\lfloor\frac{n+1}{2}\rfloor,\vspace{-0.25cm}$$ where $\lfloor\cdot\rfloor$ denotes the floor function.

\section{Consistency of the sample $d_\theta$-median}
\label{author_sec:4}
In this section we investigate the strong consistency of the sample $d_\theta$-median under general conditions.

\begin{theorem}\label{consistency}
Let $X$ be a random interval associated with a probability space $(\Omega,\mathcal A, P)$ such that the $d_\theta$-median exists and is unique. Then, the sample $d_\theta$-median is a strongly consistent estimator of the $d_\theta$-median, that is,
$$\underset{n \rightarrow \infty}{\lim}d_\theta(\widehat{\mathrm{M}_\theta[X]_n},M_\theta[X])=0 \quad \text{a.s.} [P].$$
\end{theorem}
\noindent{\emph{Proof.}}
Sufficient conditions for the strong consistency of an estimator are given in Huber~(\citeyear{huber}). We will check that these conditions, detailed below, are satisfied in our case:
\begin{itemize}
\item The parameter set ($\mathbb R \times [0,\infty)$ in our case, with the topology induced by the $d_\theta$-metric) is a locally compact space with a countable base and $(\Omega,\mathcal A,P)$ is a probability space.
\end{itemize}
Let $\rho(\omega,(y,z))$ be the following real-valued function on $\Omega\times(\mathbb R\times [0,\infty))$:
$$\begin{array}{rccl}
\rho: & \Omega \times (\mathbb R \times [0,\infty)) & \longrightarrow & \mathbb R \\
& (\omega,(y,z)) & \longmapsto & \displaystyle{d_\theta(X(\omega),[y-z,y+z])}.
\end{array}$$
\begin{itemize}
\item Assuming that $\omega_1,\omega_2 \ldots$ are independent $\Omega$-valued random elements with 

\noindent common probability distribution $P$, the sequence of functions $\{T_n\}_{n\in \mathbb N}$, defined as $T_n(\omega_1,\ldots,\omega_n)=\widehat{\mathrm{M}_\theta[(X(\omega_1),\ldots,X(\omega_n))]}_n$, satisfies that \vspace{-0.3cm}
{\small $$\frac{1}{n}\sum_{i=1}^n d_\theta(X(\omega_i),T_n(\omega_1,\ldots,\omega_n))-\inf_{(y,z) \in \mathbb R\times [0,\infty)}\frac{1}{n} \sum_{i=1}^n d_\theta (X(\omega_i),[y-z,y+z])\underset{n}{\longrightarrow} 0\vspace{-0.2cm}$$}\vspace{-0.2cm}
\noindent almost surely (obviously because of the definition of the sample $d_\theta$-median).
\end{itemize}\vspace{0.15cm}

\noindent \emph{Assumption (A-1)} For each fixed $(y_0,z_0)\in \mathbb R\times [0,\infty)$, the function \vspace{-0.1cm}
$$\begin{array}{rccll}
\rho_0: & \Omega & \longrightarrow & \mathbb R &\\
& \omega & \longmapsto & \displaystyle{\rho(\omega,(y_0,z_0))}&\displaystyle{=d_\theta(X(\omega),[y_0-z_0,y_0+z_0])}\\[0.8ex]
& & & &\displaystyle{=\sqrt{({\rm mid}\, X(\omega)-y_0)^2+\theta \cdot ({\rm spr}\, X(\omega)-z_0)^2}}\vspace{-0.1cm}
\end{array}$$
is $\mathcal A$-measurable and separable in Doob's sense: there is a P-null set $N$ and a countable subset $S\subset \mathbb R \times [0,\infty)$ such that for every open set $U\subset \mathbb R \times [0,\infty)$ and every closed interval $A$, the sets \vspace{-0.2cm}
$$V_1=\{\omega: \rho(\omega,(y,z))\in A, \forall (y,z)\in U\}\vspace{-0.1cm}$$
$$V_2=\{\omega:\rho(\omega,(y,z))\in A,\forall (y,z)\in U\cap S\}$$
differ by at most a subset of $N$.\vspace{0.15cm}

\noindent \emph{Assumption (A-2)} The function $\rho$ is a.s. lower semicontinuous in $(y_0,z_0)$, that is,
$$\underset{(y,z)\in U}{\inf}\rho(\omega,(y,z))\longrightarrow \rho(\omega,(y_0,z_0)),$$
as the neighborhood $U$ of $(y_0,z_0)$ shrinks to $\{(y_0,z_0)\}$.\vspace{0.15cm}

\noindent \emph{Assumption (A-3)} There is a measurable function $a: \Omega \rightarrow \mathbb R$ such that \vspace{-0.1cm}
$$E[\rho(\omega,(y,z))-a(\omega)]^-<\infty \quad \text{ for all } (y,z)\in \mathbb R \times [0,\infty),\vspace{-0.2cm}$$
$$E[\rho(\omega,(y,z))-a(\omega)]^+<\infty \quad \text{ for some } (y,z)\in \mathbb R \times [0,\infty).$$
Thus, $\gamma((y,z))=E[\rho(\omega,(y,z))-a(\omega)]$ is well-defined for all $(y,z)$.\vspace{0.15cm}

\noindent \emph{Assumption (A-4)} There is a $(y_0,z_0)\in \mathbb R \times [0,\infty)$ such that $\gamma ((y,z))$

\noindent $>\gamma((y_0,z_0))$ for all $(y,z)\neq (y_0,z_0).$\vspace{0.15cm}

\noindent \emph{Assumption (A-5)} There is a continuous function $b((y,z))>0$ such that
\begin{itemize}
\item for some integrable $h$, $$\underset{(y,z)\in \mathbb R\times [0,\infty)}{\inf}\frac{\rho(\omega,(w,z))-a(\omega)}{b((y,z))}\geq h(\omega).$$
\item the following condition is satisfied: $$\underset{(y,z)\rightarrow \infty}{\liminf} b((y,z))>\gamma((y_0,z_0)).$$
\item it is also fulfilled that: $$E\left[\underset{(y,z)\rightarrow \infty}{\liminf} \frac{\rho(\omega,(y,z))-a(\omega)}{b((y,z))}\right]\geq 1.$$
\end{itemize}

We now verify these conditions of Huber:\vspace{0.15cm}

\noindent \emph{(A-1)} For each fixed $(y_0,z_0)\in \mathbb R\times [0,\infty)$, the function $\rho_0$ is $\mathcal A$-measurable (because ${\rm mid}\, X$ and ${\rm }\, X$ are measurable functions since $X$ is a random interval) and separable in Doob's sense: choosing $S=\mathbb Q \times (\mathbb Q \cap [0,\infty))$ as countable subset, for every open set $U \subset \mathbb R \times [0,\infty)$ and every closed interval A, it will be seen that the sets
$$V_1=\{\omega: \rho_0(\omega)\in A, \forall (y,z)\in U\},\,\,V_2=\{\omega: \rho_0(\omega)\in A,\forall (y,z)\in U\cap S\}$$
coincide. Obviously, $V_1\subseteq V_2$. By \emph{reductio ad absurdum}, it is now supposed that $V_2\cap V_1^c\neq \emptyset$. Let $\omega_0 \in V_2\cap V_1^c$:

\begin{itemize}
\item Since $\omega_0 \in V_2$, $\rho(\omega_0,(y,z))\in A$ for all $(y,z)\in U\cap S$;\vspace{0.1cm}
\item Since $\omega_0 \in V_1^c$, there exists $(y_0,z_0)\in U$ such that $\rho(\omega_0,(y_0,z_0))\in A^c$. $A^c$ is an open set, so there exists a ball of radius $r>0$ such that $$(\rho(\omega_0,(y_0,z_0))-r,\rho(\omega_0,(y_0,z_0))+r)\subseteq A^c.$$
\end{itemize}

Notice now that, for a fixed $\omega \in \Omega$, the function
$$\begin{array}{rccll}
\rho_\omega: & \mathbb R \times [0,\infty) & \longrightarrow & \mathbb R &\\
& (y,z) & \longmapsto & \displaystyle{\rho(\omega,(y,z))}&\displaystyle{=\sqrt{({\rm mid}\, X(\omega)-y)^2+\theta \cdot ({\rm spr}\, X(\omega)-z)^2}}
\end{array}$$ is continuous. Therefore, $\rho_{\omega_0}^{-1}(\rho(\omega_0,(y_0,z_0))-r,\rho(\omega_0,(y_0,z_0))+r)$ is an open set of $\mathbb R \times [0,\infty)$ and $U\cap \rho_{\omega_0}^{-1}(\rho(\omega_0,(y_0,z_0))-r,\rho(\omega_0,(y_0,z_0))+r)\neq \emptyset$ too. $S$ is a dense set of $\mathbb R \times [0,\infty)$, so $$U\cap \rho_{\omega_0}^{-1}(\rho(\omega_0,(y_0,z_0))-r,\rho(\omega_0,(y_0,z_0))+r) \cap S \neq \emptyset.\vspace{-0.1cm}$$

\noindent Let $(y',z')\in U\cap \rho_{\omega_0}^{-1}(\rho(\omega_0,(y_0,z_0))-r,\rho(\omega_0,(y_0,z_0))+r) \cap S$. Then, $(y',z')\in U\cap S$, so $\rho(\omega_0,(y',z'))\in A$. But also, \vspace{-0.15cm} $$\rho(\omega_0,(y',z'))\in (\rho(\omega_0,(y_0,z_0))-r,\rho(\omega_0,(y_0,z_0))+r) \subset A^c.\vspace{-0.15cm}$$ This is a contradiction, so the conclusion is that $V_2\subseteq V_1$.\vspace{0.15cm}

\noindent \emph{(A-2)} Indeed, it will be proved for all $\omega \in \Omega$. Let $\omega$ be any element of $\Omega$ and let $(y_0,z_0)$ be any (fixed) point of $\mathbb R \times [0,\infty)$.

First, notice that it is fulfilled for a sequence of neighborhoods $\{U_n\}_{n\in \mathbb N}$ of $(y_0,z_0)$ when $U_n \supseteq U_{n+1}$ for all $n$ that $$\left\{\underset{(y,z)\in U_n}{\inf}d_\theta(X(\omega),[y-z,y+z])\right\}_{n\in \mathbb N}$$ is a monotonically increasing sequence. Furthermore, this sequence is bounded since
$$\underset{(y,z)\in U_n}{\inf}d_\theta(X(\omega),[y-z,y+z]) \leq d_\theta(X(\omega),[y_0-z_0,y_0+z_0])$$
for all $n\in \mathbb N$ because $\displaystyle{(y_0,z_0)\in \cap_{n\in \mathbb N}U_n}$. Therefore, the sequence converges to its supremum, which will be $ d_\theta(X(\omega),[y_0-z_0,y_0+z_0])$.

By \emph{reductio ad absurdum}, suppose that there is a smaller upper bound \vspace{-0.05cm}$$c= d_\theta(X(\omega),[y_0-z_0,y_0+z_0])-\varepsilon,$$
for an arbitrary $\varepsilon >0$. Let's denote by $U_{n_0}$ a neighborhood of $(y_0,z_0)$ satisfying that $U_{n_0}\subseteq B((y_0,z_0),\frac{\varepsilon}{2})$. Then, it can be seen that $$c < \underset{(y,z)\in U_{n_0}}{\inf}d_\theta(X(\omega),[y-z,y+z]),$$ so $c$ cannot be the supremum. Thus, using the triangular inequality,
$$\underset{(y,z)\in U_{n_0}}{\inf}d_\theta(X(\omega),[y-z,y+z])\geq \underset{(y,z)\in B((y_0,z_0),\frac{\varepsilon}{2})}{\inf}d_\theta(X(\omega),[y-z,y+z])$$ $$ \geq \underset{(y,z)\in B((y_0,z_0),\frac{\varepsilon}{2})}{\inf}\left[d_\theta(X(\omega),[y_0-z_0,y_0+z_0])-d_\theta([y-z,y+z],[y_0-z_0,y_0+z_0])\right]$$
$$=d_\theta(X(\omega),[y_0-z_0,y_0+z_0])-\underset{(y,z)\in B((y_0,z_0),\frac{\varepsilon}{2})}{\sup}d_\theta([y-z,y+z],[y_0-z_0,y_0+z_0])$$
$$> d_\theta(X(\omega),[y_0-z_0,y_0+z_0])-\varepsilon = c.$$

Now this result will be extended to general sequences $\{U_n\}_{n \in \mathbb N}$. Consider the suprema and the infima radii reached in every neighborhood, namely,
$$r_n = \underset{(y,z)\in U_n}{\sup}d_\theta([y_0-z_0,y_0+z_0],[y-z,y+z]),$$
$$s_n = \underset{(y,z)\in U_n}{\inf}d_\theta([y_0-z_0,y_0+z_0],[y-z,y+z]).$$

It is known that $r_n \underset{n}{\longrightarrow} 0$, since $\{U_n\}_{n \in \mathbb N}$ shrinks to $\{(y_0,z_0)\}$. Moreover, $s_n \underset{n}{\longrightarrow} 0$ as $0\leq s_n \leq r_n$ for all $n\in \mathbb N$.

Let $\varepsilon$ be any nonnegative number. As $r_n\underset{n}{\longrightarrow} 0$, there exists $n_1 \in \mathbb N$ such that for all $n>n_1$, $r_n<\varepsilon$. Then, $U_n \subseteq B((y_0,z_0),r_n)$ and
$$\underset{(y,z)\in U_n}{\inf}d_\theta(X(\omega),[y-z,y+z])\geq \underset{(y,z)\in B((y_0,z_0),r_n)}{\inf}d_\theta(X(\omega),[y-z,y+z])$$
$$\geq d_\theta(X(\omega),[y_0-z_0,y_0+z_0])-\underset{(y,z)\in B((y_0,z_0),r_n)}{\sup}d_\theta([y_0-z_0,y_0+z_0],[y-z,y+z])$$
$$>d_\theta(X(\omega),[y_0-z_0,y_0+z_0])-\varepsilon.$$

Analogously, as $s_n\underset{n}{\longrightarrow} 0$, there exists $n_2 \in \mathbb N$ such that for all $n>n_2$, $s_n<\varepsilon$. Therefore, $U_n \supseteq B((y_0,z_0),s_n)$ and
$$\underset{(y,z)\in U_n}{\inf}d_\theta(X(\omega),[y-z,y+z]) \leq \underset{(y,z)\in B((y_0,z_0),s_n)}{\inf}d_\theta(X(\omega),[y-z,y+z])$$
$$\leq d_\theta(X(\omega),[y_0-z_0,y_0+z_0])+\underset{(y,z)\in B((y_0,z_0),s_n)}{\inf}d_\theta([y-z,y+z],[y_0-z_0,y_0+z_0])$$
$$<d_\theta(X(\omega),[y_0-z_0,y_0+z_0])+\varepsilon.$$

So for any $\varepsilon >0$, there exists $n_0=\max\{n_1,n_2\}$, such that for all $n>n_0$,
$$d_\theta(X(\omega),[y_0-z_0,y_0+z_0])-\varepsilon < \underset{(y,z)\in U_n}{\inf}d_\theta(X(\omega),[y-z,y+z])$$ $$<d_\theta(X(\omega),[y_0-z_0,y_0+z_0])+\varepsilon,$$
that is to say,
$$\left|\underset{(y,z)\in U_n}{\inf}d_\theta(X(\omega),[y-z,y+z])-d_\theta(X(\omega),[y_0-z_0,y_0+z_0])\right|<\varepsilon,$$
so the sequence $\left\{\underset{(y,z)\in U_n}{\inf}d_\theta(X(\omega),[y-z,y+z])\right\}_{n\in \mathbb N}$ converges to

\noindent $d_\theta(X(\omega),[y_0-z_0,y_0+z_0]).$\vspace{0.15cm}

\noindent \emph{(A-3)} Let $a$ be the measurable function (see (A-1)):\vspace{-0.2cm}
$$\begin{array}{rccll}
a: & \Omega & \longrightarrow & \mathbb R &\\
& \omega & \longmapsto & \displaystyle{d_\theta(X(\omega),[0,0])=\sqrt{({\rm mid}\, X(\omega))^2+\theta \cdot ({\rm spr}\, X(\omega))^2.}}\vspace{-0.1cm}
\end{array}$$

Fixed any arbitrary $(y,z)\in \mathbb R \times [0,\infty)$,
$$E[\rho(\omega,(y,z))-a(\omega)]^-$$$$=\int_\Omega -\min \{d_\theta(X(\omega),[y-z,y+z])-d_\theta(X(\omega),[0,0]),0\}\, dP(\omega)$$
$$=\int_{\scriptsize{\begin{array}{l}\{\omega \in \Omega\,:\, d_\theta(X(\omega),[0,0])\\>d_\theta(X(\omega),[y-z,y+z])\}\end{array}}}\big[d_\theta(X(\omega),[0,0])-d_\theta(X(\omega),[y-z,y+z])\big] dP(\omega).$$
By the triangular inequality,
$$\leq \int_{\scriptsize{\begin{array}{l}\{\omega \in \Omega\,:\, d_\theta(X(\omega),[0,0])\\>d_\theta(X(\omega),[y-z,y+z])\}\end{array}}}\big[d_\theta(X(\omega),[y-z,y+z])+d_\theta([y-z,y+z],[0,0])$$$$-d_\theta(X(\omega),[y-z,y+z])\big] dP(\omega)$$
$$= d_\theta([y-z,y+z],[0,0])\cdot P\big(\omega : d_\theta(X(\omega),[0,0])>d_\theta(X(\omega),[y-z,y+z])\big) < \infty.$$

Analogously,
$$E[\rho(\omega,(y,z))-a(\omega)]^+$$$$=\int_\Omega \max \{d_\theta(X(\omega),[y-z,y+z])-d_\theta(X(\omega),[0,0]),0\}\, dP(\omega)$$
$$=\int_{\scriptsize{\begin{array}{l}\{\omega \in \Omega\,:\, d_\theta(X(\omega),[0,0])\\\leq d_\theta(X(\omega),[y-z,y+z])\}\end{array}}}\big[d_\theta(X(\omega),[y-z,y+z])-d_\theta(X(\omega),[0,0])\big] \, dP(\omega).$$
By the triangular inequality,
$$\leq \int_{\scriptsize{\begin{array}{l}\{\omega \in \Omega\,:\, d_\theta(X(\omega),[0,0])\\ \leq d_\theta(X(\omega),[y-z,y+z])\}\end{array}}}\big[d_\theta(X(\omega),[0,0])+d_\theta([0,0],[y-z,y+z])$$$$-d_\theta(X(\omega),[0,0])\big] dP(\omega)$$
$$= d_\theta([0,0],[y-z,y+z])\cdot P\big(\omega : d_\theta(X(\omega),[0,0])\leq d_\theta(X(\omega),[y-z,y+z])\big) < \infty.$$

So the second inequality also holds for all $(y,z)\in \mathbb R\times [0,\infty)$ in this case.\vspace{0.15cm}

\noindent \emph{(A-4)} The $d_\theta$-median exists and is unique, so that \vspace{-0.1cm}
$$(\hbox {\rm mid}\, M_\theta[X], \hbox {\rm spr}\, M_\theta[X])=\arg \underset{(y,z)\in \mathbb R\times [0,\infty)}{\min}E\left[d_\theta(X(\omega),[y-z,y+z])\right]$$
$$=\arg \underset{(y,z)\in \mathbb R\times [0,\infty)}{\min} E\left[d_\theta(X(\omega),[y-z,y+z])\right]-E\left[d_\theta(X(\omega),[0,0])\right]$$$$=\arg \underset{(y,z)\in \mathbb R\times [0,\infty)}{\min} E\left[d_\theta(X(\omega),[y-z,y+z])-d_\theta(X(\omega),[0,0])\right]$$$$=\arg \underset{(y,z)\in \mathbb R\times [0,\infty)}{\min}\gamma((y,z))$$\vspace{0.1cm}
and $(y_0,z_0):=({\hbox {\rm mid}}\, M_\theta[X], {\hbox {\rm spr}}\, M_\theta[X])$ fulfills this assumption.\vspace{0.15cm}

\noindent \emph{(A-5)} There is a continuous function $b((y,z))>0$ \vspace{-0.1cm}
$$\begin{array}{rccll}
b: & \mathbb R\times [0,\infty) & \longrightarrow & \mathbb R &\\
& (y,z) & \longmapsto & \displaystyle{d_\theta([y-z,y+z],[0,0])+1}\vspace{-0.1cm}
\end{array}$$
such that
\begin{itemize}
\item for the integrable function $h(\omega):=-1$, $$\underset{(y,z)\in \mathbb R\times [0,\infty)}{\inf}\frac{d_\theta(X(\omega),[y-z,y+z])-d_\theta(X(\omega),[0,0])}{d_\theta([y-z,y+z],[0,0])+1}\geq -1$$
because using the triangular inequality,
$$\underset{(y,z)\in \mathbb R\times [0,\infty)}{\inf}\frac{d_\theta(X(\omega),[y-z,y+z])-d_\theta(X(\omega),[0,0])}{d_\theta([y-z,y+z],[0,0])+1}$$
$$\geq \underset{(y,z)\in \mathbb R\times [0,\infty)}{\inf}\frac{d_\theta(X(\omega),[0,0])-d_\theta([y-z,y+z],[0,0])-d_\theta(X(\omega),[0,0])}{d_\theta([y-z,y+z],[0,0])+1}$$
$$=\underset{(y,z)\in \mathbb R\times [0,\infty)}{\inf}\frac{-d_\theta([y-z,y+z],[0,0])}{d_\theta([y-z,y+z],[0,0])+1}\geq -1.$$\vspace{0.15cm}
\item the following condition is satisfied: $$\underset{(y,z)\rightarrow \infty}{\liminf} b((y,z))>\gamma((y_0,z_0)).$$\vspace{0.15cm}
Let $\{(y_n,z_n)\}\subset \mathbb R\times [0,\infty)$ be any sequence with $(y_n,z_n)\underset{n}{\longrightarrow}\infty$ (i.e., $d_\theta([y_n-z_n,y_n+z_n],[0,0])\underset{n}{\longrightarrow}\infty$) and $$M=E\left[d_\theta(X(\omega),[y_0-z_0,y_0+z_0])-d_\theta(X(\omega),[0,0])\right]=\gamma((y_0,z_0))\in \mathbb R,$$ where $(y_0,z_0)$ represents the minimum found in (A-4). Then, there exists $n_0\in \mathbb N$ such that for all $n\geq n_0$,
$$d_\theta([y_n-z_n,y_n+z_n],[0,0])>M.$$
So, for all $n\geq n_0$,
$$\underset{k\geq n}{\inf}b((y_k,z_k))=\underset{k\geq n}{\inf}\left(d_\theta([y_k-z_k,y_k+z_k],[0,0])+1\right)\geq M+1.$$
Finally,
$$\underset{n\rightarrow \infty}{\liminf}b((y_n,z_n))=\underset{n\rightarrow \infty}{\lim}(\underset{k\geq n}{\inf} b((y_k,z_k)))\geq M+1>M=\gamma((y_0,z_0)).$$
\item it is also fulfilled that: $$E\left[\underset{(y,z)\rightarrow \infty}{\liminf} \frac{d_\theta(X(\omega),[y-z,y+z])-d_\theta(X(\omega),[0,0])}{b((y,z))}\right]\geq 1.$$
Let's see that $$\underset{(y,z)\rightarrow \infty}{\liminf} \frac{d_\theta(X(\omega),[y-z,y+z])-d_\theta(X(\omega),[0,0])}{d_\theta([y-z,y+z],[0,0])+1}\geq 1,$$
so the result follows.
$$\underset{(y,z)\rightarrow \infty}{\liminf} \frac{d_\theta(X(\omega),[y-z,y+z])-d_\theta(X(\omega),[0,0])}{d_\theta([y-z,y+z],[0,0])+1}$$
$$=\underset{n\rightarrow \infty}{\lim}\left(\underset{k\geq n}{\inf} \frac{d_\theta(X(\omega),[y_k-z_k,y_k+z_k])-d_\theta(X(\omega),[0,0])}{d_\theta([y_k-z_k,y_k+z_k],[0,0])+1}\right)$$ for any fixed $\omega\in\Omega$. The sequence $$\left\{\underset{k\geq n}{\inf} \frac{d_\theta(X(\omega),[y_k-z_k,y_k+z_k])-d_\theta(X(\omega),[0,0])}{d_\theta([y_k-z_k,y_k+z_k],[0,0])+1}\right\}_{n\in\mathbb N}$$ is monotonically increasing and is upper bounded by $1$: for all $k\in \mathbb N$, using the triangular inequality,
$$\frac{d_\theta(X(\omega),[y_k-z_k,y_k+z_k])-d_\theta(X(\omega),[0,0])}{d_\theta([y_k-z_k,y_k+z_k],[0,0])+1}$$ $$\leq \frac{d_\theta([y_k-z_k,y_k+z_k],[0,0])}{d_\theta([y_k-z_k,y_k+z_k],[0,0])+1}\leq 1.$$

So it converges to its supremum:
$$\underset{n\rightarrow \infty}{\lim}\left(\underset{k\geq n}{\inf} \frac{d_\theta(X(\omega),[y_k-z_k,y_k+z_k])-d_\theta(X(\omega),[0,0])}{d_\theta([y_k-z_k,y_k+z_k],[0,0])+1}\right)$$
$$=\underset{n}{\sup}\left(\underset{k\geq n}{\inf} \frac{d_\theta(X(\omega),[y_k-z_k,y_k+z_k])-d_\theta(X(\omega),[0,0])}{d_\theta([y_k-z_k,y_k+z_k],[0,0])+1}\right)$$
Let's finally see that this supremum is at least equal to $1$. By \emph{reductio ad absurdum}, let's suppose that
$$\underset{n}{\sup}\left(\underset{k\geq n}{\inf} \frac{d_\theta(X(\omega),[y_k-z_k,y_k+z_k])-d_\theta(X(\omega),[0,0])}{d_\theta([y_k-z_k,y_k+z_k],[0,0])+1}\right)=1-\varepsilon,$$
for some $\varepsilon >0$. One gets then a contradiction because one finds an $n^*\in \mathbb N$ such that
$$\underset{k\geq n^*}{\inf} \frac{d_\theta(X(\omega),[y_k-z_k,y_k+z_k])-d_\theta(X(\omega),[0,0])}{d_\theta([y_k-z_k,y_k+z_k],[0,0])+1}>1-\varepsilon$$
since for all $k\geq n^*$, $$\frac{d_\theta(X(\omega),[y_k-z_k,y_k+z_k])-d_\theta(X(\omega),[0,0])}{d_\theta([y_k-z_k,y_k+z_k],[0,0])+1}\geq 1-\frac{\varepsilon}{2}>1-\varepsilon$$
as we will show now. Recall that $(y_n,z_n)\underset{n}{\longrightarrow}\infty$, so for all $M\in \mathbb R$, there exists $n^*\in \mathbb N$ such that for all $n\geq n^*$, $d_\theta([y_n-z_n,y_n+z_n],[0,0])>M$. Therefore,
$$d_\theta([y_n-z_n,y_n+z_n],X(\omega))\geq d_\theta([y_n-z_n,y_n+z_n],[0,0])-d_\theta(X(\omega),[0,0])$$$$>M-d_\theta(X(\omega),[0,0]).$$
Taking $M:=\frac{2}{\varepsilon}-1+\frac{4}{\varepsilon}\cdot d_\theta(X(\omega),[0,0])\in \mathbb R$ (for the fixed arbitrary $\omega\in\Omega$), we can easily check that $1-\frac{\varepsilon}{2}$ is a lower bound of the sequence $$\left\{ \frac{d_\theta(X(\omega),[y_k-z_k,y_k+z_k])-d_\theta(X(\omega),[0,0])}{d_\theta([y_k-z_k,y_k+z_k],[0,0])+1}\right\}_{k\geq n^*}.$$

For any $k\geq n^*$,
$$d_\theta(X(\omega),[y_k-z_k,y_k+z_k])-d_\theta(X(\omega),[0,0])$$
$$=\left(1-\frac{\varepsilon}{2}\right)d_\theta(X(\omega),[y_k-z_k,y_k+z_k])+\frac{\varepsilon}{2}d_\theta(X(\omega),[y_k-z_k,y_k+z_k])$$
$$-d_\theta(X(\omega),[0,0])$$
$$\geq \left(1-\frac{\varepsilon}{2}\right)d_\theta([y_k-z_k,y_k+z_k],[0,0])-\left(1-\frac{\varepsilon}{2}\right)d_\theta(X(\omega),[0,0])$$
$$+\frac{\varepsilon}{2}d_\theta(X(\omega),[y_k-z_k,y_k+z_k])-d_\theta(X(\omega),[0,0])$$
$$=\left(1-\frac{\varepsilon}{2}\right)d_\theta([y_k-z_k,y_k+z_k],[0,0])+\frac{\varepsilon}{2}d_\theta(X(\omega),[y_k-z_k,y_k+z_k])$$
$$-\left(2-\frac{\varepsilon}{2}\right)d_\theta(X(\omega),[0,0])$$
$$>\left(1-\frac{\varepsilon}{2}\right)d_\theta([y_k-z_k,y_k+z_k],[0,0])+\frac{\varepsilon}{2}\left(\frac{2}{\varepsilon}-1+\Big(\frac{4}{\varepsilon}-1\Big)d_\theta(X(\omega),[0,0])\right)$$
$$-\left(2-\frac{\varepsilon}{2}\right)d_\theta(X(\omega),[0,0])=\left(1-\frac{\varepsilon}{2}\right)d_\theta([y_k-z_k,y_k+z_k],[0,0])+1-\frac{\varepsilon}{2}$$
$$\hspace{1.9cm}=\left(1-\frac{\varepsilon}{2}\right)\big(d_\theta([y_k-z_k,y_k+z_k],[0,0])+1\big).\hspace{2.4cm}\square$$
\end{itemize}

\section{Concluding remarks}

\label{author_sec:8}
This paper complements the study of the properties of the 
$d_{\theta}$-median as a robust estimator of the center of a random interval by showing its strong consistency which is one of the most important basic properties of an estimator. We obtained this result by showing that all the sufficient conditions of Huber (\citeyear{huber})
are fulfilled. These results open the door to further develop robust statistical inference for random 
intervals based on the $d_{\theta}$-median such as the development of hypotheses testing procedures.

\begin{acknowledgements}
Authors are grateful to Mar\'ia \'Angeles Gil for her helpful suggestions to improve this paper. The research by Beatriz Sinova was partially supported by/benefited from the Spanish Ministry of Science and Innovation Grant MTM2009-09440-C02-01. She has been also granted with the Ayuda del Programa de FPU AP2009-1197 from the Spanish Ministry of Education and the Ayuda para Estancias Breves del Programa FPU EST12/00344, an Ayuda de Investigaci\'on 2011 from the Fundaci\'on Banco Herrero and three Short Term Scientific Missions associated with the COST Action IC0702. The research by Stefan Van Aelst was supported by a grant of the Fund for Scientific Research-Flanders (FWO-Vlaanderen) and by IAP research network grant nr. P7/06 of the Belgian government (Belgian Science Policy). Their financial support is gratefully acknowledged.
\end{acknowledgements}


\begin{thebibliography}{}{\footnotesize
%
%

\bibitem[(2014)]{Arribas2014} Arribas-Gil A, M\"{u}ller H-G (2014) Pairwise dynamic time warping for event data. Comput Stat Data Anal 69:255--268
\bibitem[(1965)]{Aumann1965} Aumann RJ (1965) Integrals of set-valued functions. J Math Anal Appl 12:1--12
\bibitem[(1995)]{bertoluzza1995} Bertoluzza C, Corral N, Salas A (1995) On a new class of distances between fuzzy numbers. Math \& Soft Comput 2:71--84
\bibitem[(2013)]{bianco} Bianco AM, Boente G, Rodrigues IM (2013) Robust tests in generalized linear models with missing responses. Comput Stat Data Anal 65:80--97
\bibitem[(2003)]{Billard2003} Billard L, Diday E (2003) From the Statistics of data to the Statistics of knowledge: Symbolic Data Analysis. J Am Stat Ass 98:470--487
\bibitem[(2011)]{Blanco2011} Blanco-Fern\'{a}ndez A, Corral N, Gonz\'{a}lez-Rodr\'{\i}guez G (2011) Estimation of a flexible simple linear model for interval data based on set arithmetic. Comput Stat Data Anal 55(9):2568--2578
\bibitem[(2013)]{Blanco2013} Blanco-Fern\'{a}ndez A, Colubi A, Garc\'{\i}a-B\'{a}rzana M (2013) A set arithmetic-based linear regression model for modelling interval-valued responses through real-valued variables. Inform Sci 247:109--122
\bibitem[(2012)]{Brito2012} Brito P, Duarte Silva AP (2012) Modelling interval data with Normal and Skew-Normal distributions. J Appl Stat 39(1):3--20
\bibitem[(1983)]{Brown} Brown BM (1983) Statistical uses of the spatial median. J Royal Stat Soc Ser B 45(1):25--30
\bibitem[(2001)]{cadre2001} Cadre B (2001) Convergent estimators for the $L_1$-median of a Banach valued random variable. Statistics 35:509--521
\bibitem[(2006)]{Carvalho2006} De Carvalho FDAT, Brito P, Bock HH (2006) Dynamic clustering for interval data based on $L_2$ distance. Comput Stat 21(2):231--250
\bibitem[(2012)]{Coppi2012} Coppi R, D'Urso P, Giordani P (2012) Fuzzy and possibilistic clustering for fuzzy data. Comput Stat Data Anal 56(4):915--927
\bibitem[(2013)]{DaCosta2013} Da Costa AFBF, Pimentel BA, De Souza RMCR (2013) Clustering interval data through kernel-induced feature space. J Intell Inf Syst 40(1):101--140
\bibitem[(2010)]{Debruyne} Debruyne M, Hubert M, Van Horebeek J (2010) Detecting influential observations in Kernel PCA. Comput Stat Data Anal 54(12):3007--3019
\bibitem[(1983)]{donoho1983} Donoho DL, Huber PJ (1983) The notion of breakdown point. In: Bickel PJ, Doksum K, Hodges Jr JL (Eds.) A Festschrift for Erich L. Lehmann. Wadsworth, Belmont.
\bibitem[(1987)]{ducharme1987} Milasevic P, Ducharme GR (1987) Uniqueness of the spatial median. Ann Statist 15:1332--1333
\bibitem[(2004)]{DUrso2004} D'Urso P, Giordani P (2004) A least squares approach to principal component analysis for interval valued data. Chemometr Intell Lab 70(2):179--192
\bibitem[(2006)]{DUrso2006} D'Urso P, Giordani P (2006) A robust fuzzy k-means clustering model for interval valued data. Comput Stat 21(2):251--269
\bibitem[(2011)]{DUrso2011} D'Urso P, De Giovanni L (2011) Midpoint radius self-organizing maps for interval-valued data with telecommunications application. Appl Soft Comput 11(5):3877--3886
\bibitem[(2014)]{DUrso2014} D'Urso P, De Giovanni L, Massari R (2014) Trimmed fuzzy clustering for interval-valued data. Adv Data Anal Classif, in press (doi:10.1007/s11634-014-0169-3)
\bibitem[(2013)]{fagundes2013} Fagundes RAA, De Souza RMCR, Cysneiros FJA (2013) Robust regression with application to symbolic interval data. Eng Appl Art Intel 26(1):564--573
\bibitem[(2013)]{Febrero2013} Febrero-Bande M, Gonz\'{a}lez-Manteiga W (2013) Generalized additive models for functional data. Test 22(2):278--292
\bibitem[(2013)]{Ferraro2013} Ferraro MB, Giordani P (2013) On possibilistic clustering with repulsion constraints for imprecise data. Inform Sci 245:63--75
\bibitem[(2013)]{ferraty} Ferraty F, Sued M, Vieu P (2013) Mean estimation with data missing at random for functional covariables. Statistics 47(4):688--706
\bibitem[(2012)]{Fritz} Fritz H, Filzmoser P, Croux C (2012) A comparison of algorithms for the multivariate $L_1$-median. Comput Stat 27(3):393--410
\bibitem[(2002)]{gil2002} Gil MA, Lubiano MA, Montenegro M, L\'opez-Garc\'ia MT (2002) Least squares fitting of an affine function and strength of association for interval data. Metrika 56:97--111
\bibitem[(2007)]{Gil2007} Gil MA, Gonz\'alez-Rodr\'{\i}guez G, Colubi A, Montenegro M (2007) Testing linear independence in linear models with interval-valued data. Comput Stat Data Anal 51(6):3002--3015
\bibitem[(2014)]{Giordani2014} Giordani P (2014) Lasso-constrained regression analysis for interval-valued data. Adv Data Anal Classif, in press (doi:10.1007/s11634-014-0164-8)
\bibitem[(2008)]{Giusti2008} Giusti A, Grassini L (2008) Cluster analysis of census data using the symbolic data approach. Adv Data Anal Classif 2(2):163--176
\bibitem[(2007)]{Gonzalez2007} Gonz\'{a}lez-Rodr\'{\i}guez G, Blanco A, Corral N, Colubi A (2007) Least squares estimation of linear regression models for convex compact random sets. Adv Data Anal Classif 1(1):67--81
\bibitem[(2012)]{Gonzalez2012} Gonz\'{a}lez-Rodr\'{\i}guez G, Colubi A, Gil MA (2012) Fuzzy data treated as functional data: A one-way ANOVA test approach. Comput Stat Data Anal 56:943--955
\bibitem[(1974)]{gower1974} Gower JC (1974) Algorithm AS 78: The mediancentre. Appl Statist 23:466--470
\bibitem[(1967)]{huber} Huber PJ (1967) The behavior of maximum likelihood estimates under nonstandard conditions. In: Proceedings of the Fifth Berkeley Symposium on Mathematical Statistics and Probability 1, pp. 221--233
\bibitem[(1990)]{ishibuchi1990} Ishibuchi H, Tanaka H (1990) Multiobjective programming in optimization of the interval objective function. Europ J Oper Res 48:219--225
\bibitem[(2014)]{jacques} Jacques J, Preda C (2014) Model-based clustering for multivariate functional data. Comput Stat Data Anal 71:92--106
\bibitem[(2011)]{Lima2011} Lima Neto EA, Cordeiro GM, de Carvalho FAT (2011) Bivariate symbolic regression models for interval-valued variables.
J Stat Comput Simul 81(11):1727--1744
\bibitem[(2014)]{lin} Lin TH (2014) Model selection information criteria in latent class models with missing data and contingency question. J Stat Comput Simul 84(1):159--170
\bibitem[(1990)]{liu1990} Liu RY (1990). On a notion of data depth based on random simplices. Ann Statist 18:405--414
\bibitem[(2012)]{Makosso2012} Makosso-Kallyth S, Diday E (2012) Adaptation of interval PCA to symbolic histogram variables. Adv Data Anal Classif 6(2):147--159
\bibitem[(2008)]{Montenegro2008} Montenegro M, Casals MR, Colubi A, Gil MA (2008) Testing 'two-sided' hypothesis about the mean of an interval-valued random set. In: Dubois D, Lubiano MA, Prade H, Gil MA, Grzegorzewski P, Hryniewicz O (Eds.) Soft Methods for Handling Variability and Imprecision. Springer, Heidelberg, pp. 133--139
\bibitem[(2010)]{Nakama2010} Nakama T, Colubi A, Lubiano MA (2010) Two-way analysis of variance for interval-valued data. In: Borgelt C, Gonz\'{a}lez-Rodr\'{\i}guez G, Trutschnig W, Lubiano MA, Gil MA, Grzegorzewski P, Hryniewicz O (Eds.) Combining Soft Computing and Statistical Methods in Data Analysis. Springer, Heidelberg, pp. 475--482
\bibitem[(2007)]{Roelant} Roelant E, Van Aelst S (2007) An L1-type estimator of multivariate location. Stat Meth \& Appl 15:381--393
\bibitem[(1996)]{rousseeuw1996} Rousseeuw PJ, Ruts I (1996) Bivariate location depth. J. Roy. Statist. Soc. Ser. C 45:516--526
\bibitem[(1998)]{rousseeuw1998} Rousseeuw PJ, Ruts I (1998) Constructing the bivariate Tukey median. Statistica Sinica 8:827--839
\bibitem[(2012)]{sinova2012} Sinova B, Gil MA, Colubi A, Van Aelst S (2012) The median of a random fuzzy number. The 1-norm distance approach. Fuzzy Sets Syst 200:99--115
\bibitem[(2013)]{sinova2013} Sinova B, Gonz\'alez-Rodr\'iguez G, Van Aelst S (2013) An alternative approach to the median of a random interval using an $L^2$ metric. In: Kruse R, Berthold MR, Moewes C, Gil MA, Grzegorzewski P, Hryniewicz O (Eds.) Sinergies of Soft Computing and Statistics for Intelligent Data Analysis. Springer, Heidelberg, pp. 273--281
\bibitem[(2014)]{SunRalescu2014} Sun Y, Ralescu DA (2014) A normal hierarchical model and minimum contrast estimation for random intervals. Ann Inst Stat Math, in press (doi:10.1007/s10463-014-0453-1)
\bibitem[(2009)]{trutschnig2009} Trutschnig W, Gonz\'alez-Rodr\'iguez G, Colubi A, Gil MA (2009) A new family of metrics for compact, convex (fuzzy) sets based on a generalized concept of mids and spread. Inf Sci 179:3964--3972
\bibitem[(1975)]{tukey1975} Tukey JW (1975) Mathematics and the picturing of data. In: Proc. International Congress of Mathematicians, Vancouver, 1994, 2, pp. 523--531
\bibitem[(1985)]{vitale1985} Vitale RA (1985) $L_p$ metrics for compact, convex sets. J Approx Theory 45:280--287
\bibitem[(1909)]{Weber1909} Weber A (1909) \"{U}ber den Standort der Industrien. Mohr, T\"{u}bingen.
\bibitem[(2013)]{zhao} Zhao P-Y, Tang M-L, Tang N-S (2013) Robust estimation of distribution functions and quantiles with non-ignorable missing data. Canad J Stat 41(4):575--595
\bibitem[(2013)]{Zuo} Zuo Y (2013) Multidimensional medians and uniqueness.  Comput Stat Data Anal 66:82--88}


\end{thebibliography}


\end{document}